\input{aipcheck}

\documentclass[,final]
  {aipproc}

\layoutstyle{6x9}

\newcommand{\spur}[1]{\not\! #1 \,}

\begin{document}

\title{Probing Universal Extra Dimensions through rare decays induced by $b \to s$ transition}


\author{Rossella Ferrandes}{
  address={Physics Department, University of Bari and INFN, Sezione di Bari, Italy}
}

\begin{abstract}
 A few $B_{d,s}$ and $\Lambda_b$ decays induced by $b \to s$ transition
 are studied in the Standard Model and in the framework of the Appelquist, Cheng and Dobrescu (ACD) model, which is a New Physics scenario where a single
 universal extra dimension is considered. In particular, we investigate the
 sensitivity of the observables to the radius $R$ of the
 compactified extra dimension.
\end{abstract}

\maketitle

\section{Introduction}

 \noindent Among the ideas proposed to extend the SM, a lot of attention
  has recently been devoted to models including extra dimensions \cite{rev}.
 An interesting model is that proposed by Appelquist, Cheng and
Dobrescu with so-called universal extra dimensions (UED)
\cite{ACD}, which means that all the SM fields may propagate in
one or more compact extra dimensions. The compactification of the
extra dimensions involves the appearance of an infinite discrete
set of four dimensional fields which create the so-called KK
particles, the masses of which are related to compactification
radius according to the relation $ m_n^2=m_0^2+\frac{n^2}{R^2} $,
with $n=1, 2,...$.
 \noindent The simplest UED scenario is
characterized by a single extra dimension. It presents the
remarkable feature of having only one new parameter with respect
to SM, the radius $R$ of the compactified extra dimension.

\noindent Flavor changing neutral current processes are of
particular interest, since they are sensitive to loop
contributions involving KK states and therefore can be used to
constrain their masses and couplings, i. e. the compactification
radius \cite{Agashe:2001xt}. This observation led Buras and
collaborators to compute in the ACD model the effective
Hamiltonian of several FCNC processes in particular in the $b$
sector, namely $B_{d,s}$ mixing and $b \to s$ transitions such as
$b \to s \gamma$ and $b \to s l^+ l^-$ \cite{Buras:2002ej}. In
particular, it was found that   ${\cal B}(B \to X_s \gamma)$
allowed to constrain ${1/ R} \ge 250$ GeV, a bound updated by  a
more recent analysis to $1/R \ge 600$ GeV at 95$\%$ CL, or to $1/R
\ge 330 $ GeV at 99$\%$ CL  \cite{Haisch:2007vb}.

\noindent In \cite{noi} we have analyzed several $B_{d,s}$ and
$\Lambda_b$ decays induced by $b \to s$ transitions, finding that
in many cases the hadronic uncertainties do not hide the
dependence of the observables on $R$, as discussed in the
following Sections for the decay modes $B_d \to K^* \gamma$, $B_d
\to K^* \nu \bar{\nu}$, $B_s \to \phi \gamma$, $B_s \to \phi \nu
\bar{\nu}$, $\Lambda_b \to \Lambda \gamma$ and $\Lambda_b \to
\Lambda \nu \bar{\nu}$.

\section{The decays $B \to K^* \gamma, K^* \nu \bar{\nu}$ and $B_s \to \phi \gamma, \phi \nu \bar{\nu}$}

In the Standard Model   the  $b \to s \gamma$ and  $b \to s \nu
\bar \nu$  transitions are described by the effective $\Delta
B=-1$, $\Delta S=1$ Hamiltonians

 \begin{eqnarray} H_{b \to s \gamma}=4\,{G_F
\over \sqrt{2}} V_{tb} V_{ts}^*  c_7 O_7, \hspace{0.8cm} H_{b \to
s\nu \bar \nu}= {G_F \over \sqrt{2}} {\alpha \over 2 \pi
\sin^2(\theta_W)} V_{tb} V_{ts}^* \eta_X X(x_t) \, O_L  = c_L O_L
\hspace{0.1cm} ,\label{hamilnu}
\end{eqnarray}
involving the operators
 \begin{eqnarray}
O_7 = {e \over 16 \pi^2} \big[ m_b ({\bar s}_{L } \sigma^{\mu \nu}
b_{R }) +m_s ({\bar s}_{R } \sigma^{\mu \nu} b_{L }) \big]F_{\mu
\nu}, \hspace{0.5cm} O_L = {\bar s}\gamma^\mu (1-\gamma_5) b {\bar
\nu}\gamma_\mu (1-\gamma_5) \nu \,\,. \nonumber
\end{eqnarray}
\noindent   $G_F$ is the Fermi constant and $V_{ij}$ are elements
of the CKM mixing matrix; moreover, $\displaystyle b_{R,L}={1 \pm
\gamma_5 \over 2}b$, $\alpha=\displaystyle{e^2 \over 4 \pi}$ is
the electromagnetic constant, $\theta_W$  the Weinberg angle and
$F_{\mu \nu}$ denotes the electromagnetic field strength tensor.
The  function $X(x_t)$ ($x_t=\displaystyle{ m_t^2 \over M_W^2}$,
with $m_t$  the top quark mass) has been computed in
\cite{inami,buchalla}; the QCD factor $\eta_X$ is close to one, so
that we can put  it to unity \cite{buchalla,Buchalla:1998ba}.

\noindent The effect of the Kaluza-Klein states only consists in a
 modification of the Wilson coefficients $c_7$ and $c_L$ in
 (\ref{hamilnu}), which acquire a dependence
 on the compactification radius \textit{R}.
 In particular, the coefficients can be expressed in terms of
 functions $F(x_t,1/R)$, which generalize their SM analogues
 $F_0(x_t)$ according to: $ F(x_t,1/R)=F_0(x_t)+\sum_{n=1}^{\infty}F_n(x_t,x_n)$,
 with $x_n=\frac{m_n^2}{M_W^2}$, $m_n=\frac{n}{R}$.

The description of the decay modes $B \to K^* \gamma$ and  $B \to
K^* \nu \bar \nu$  involves the hadronic matrix elements of the
operators appearing in the effective Hamiltonians (\ref{hamilnu}),
which can be parameterized in the following way:
\begin{eqnarray}
<K^*(p^\prime,\epsilon)|{\bar s} \sigma_{\mu \nu} q^\nu
{(1+\gamma_5) \over 2} b |B(p)> = i \epsilon_{\mu \nu \alpha
\beta} \epsilon^{* \nu} p^\alpha p^{\prime \beta} \; 2 \;
T_1(q^2)+ \nonumber \\ +   \Big[ \epsilon^*_\mu (M_{B}^2 -
M^2_{K^*}) - \epsilon^* \cdot q (p+p')_\mu \Big] \; T_2(q^2) +
\epsilon^* \cdot q \left [ q_\mu - {q^2 \over M_{B}^2 - M^2_{K^*}}
(p + p')_\mu \right ] \; T_3(q^2),   \nonumber
\end{eqnarray}
\begin{eqnarray}
<K^*(p^\prime,\epsilon)|{\bar s} \gamma_\mu (1-\gamma_5) b |B(p)>
= \epsilon_{\mu \nu \alpha \beta} \epsilon^{* \nu} p^\alpha
p^{\prime \beta} { 2 V(q^2) \over M_{B} + M_{K^*}}
- i \Big [ \epsilon^*_\mu (M_{B} + M_{K^*}) A_1(q^2)   \nonumber \\
- (\epsilon^* \cdot q) (p+p')_\mu  {A_2(q^2) \over M_{B} + M_{K^*}
} -  (\epsilon^* \cdot q) {2 M_{K^*} \over q^2} \big(A_3(q^2) -
A_0(q^2)\big) q_\mu \Big], \nonumber
\end{eqnarray}
\noindent where $q=p-p^\prime$, and $\epsilon$ is the $K^*$ meson
polarization vector.  At zero value of $q^2$ the condition
$T_1(0)=T_2(0)$ holds, so that the $B \to K^* \gamma$ decay
amplitude involves a single hadronic parameter, $T_1(0)$.
Furthermore, a relation holds the form factors $A_1$, $A_2$ and
$A_3$: \noindent $ A_3(q^2) = {M_{B} + M_{K^*} \over 2 M_{K^*}}
A_1(q^2) - {M_{B} - M_{K^*} \over 2 M_{K^*}} A_2(q^2) $ together
with  $ A_3(0) = A_0(0)$.

\noindent We use for the form factors
 two sets of results:  the first one,  denoted as set A,
obtained by  three-point QCD sum rules based on the short-distance
expansion \cite{Colangelo:1995jv};  the second one, denoted as set
B,   obtained  by  QCD sum rules based on the light-cone expansion
\cite{Ball:2004rg}.

Let us consider the radiative mode   $ B \to  K^* \gamma$. Its
 decay rate is given by: \begin{equation} \Gamma(B \to K^* \gamma)={\alpha
G_F^2 \over 8 \pi^4}|V_{tb}V_{ts}^*|^2m_b^2 |c_7|^2 [T_1(0)]^2
M_B^3 \left( 1-{M_{K^*}^2 \over M_B^2} \right)^3 \,\,\, .
\label{ratebkstargamma} \end{equation}

\begin{figure}[ht]
\includegraphics[width=0.35\textwidth] {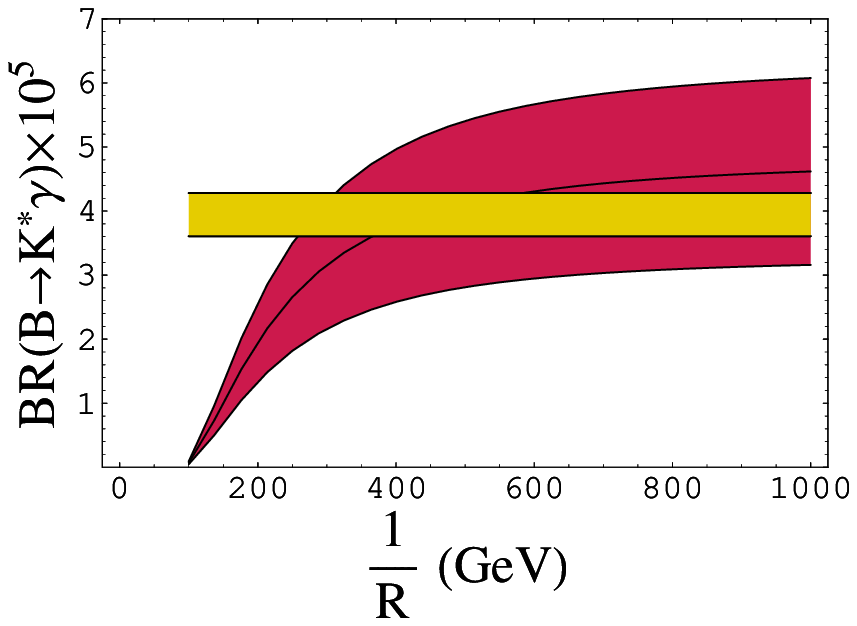}
\hspace{0.5cm}
 \includegraphics[width=0.35\textwidth] {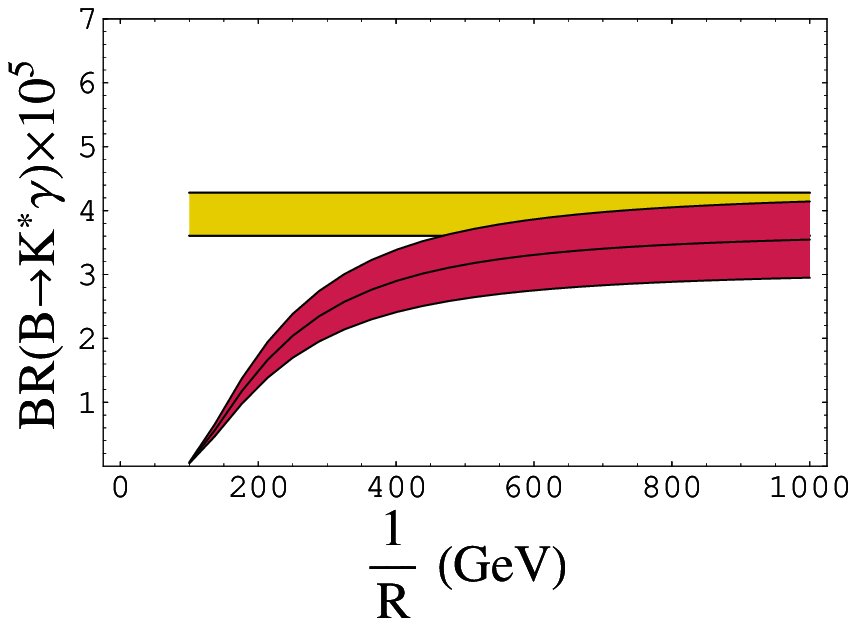}
\caption{\baselineskip=15pt ${\cal B}_{B \to K^* \gamma}$  versus
$\displaystyle{1 / R}$ using  set A  (left) and   B
 (right) of form factors .  The  horizontal
band corresponds to the experimental result. } \vspace*{1.0cm}
\label{brkstargamma}
\end{figure}

\noindent One can appreciate the consequences of the existence  of
a single universal extra dimension considering
Fig.~\ref{brkstargamma}, where the branching fraction is plotted
as a function of  $1/R$. The hadronic uncertainty is evaluated
comparing the two set of form factors and taking into account
their errors. A comparison between experimental data
\cite{Nakao:2004th} (represented by the horizontal band in the
Fig.~\ref{brkstargamma}) and theoretical predictions allows to put
a lower bound of
  $1/R \ge 250$ GeV  adopting set A, and a stronger bound of $1/R \ge
400$ GeV for set B.

\begin{figure}[ht]
\includegraphics[width=0.35\textwidth] {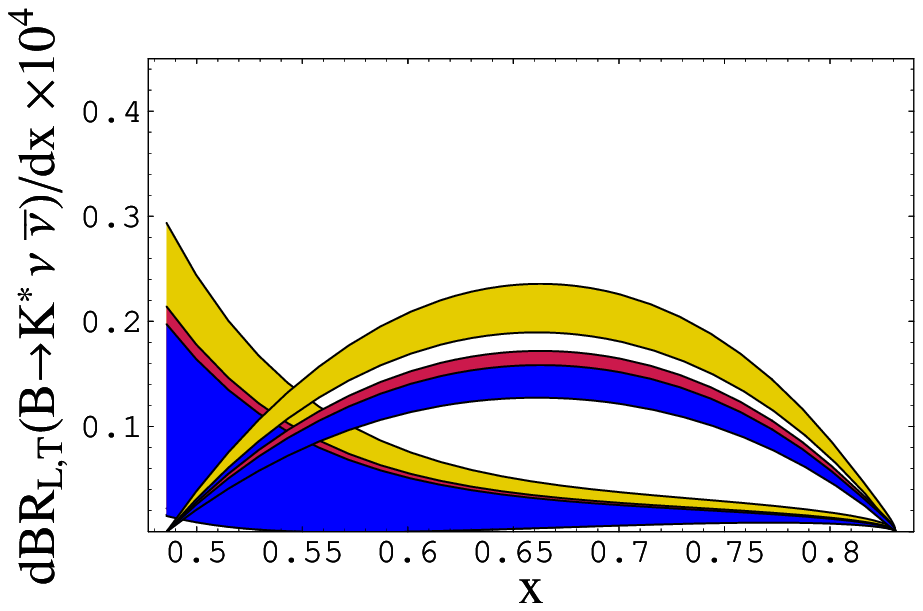} \hspace{0.5cm}
 \includegraphics[width=0.35\textwidth] {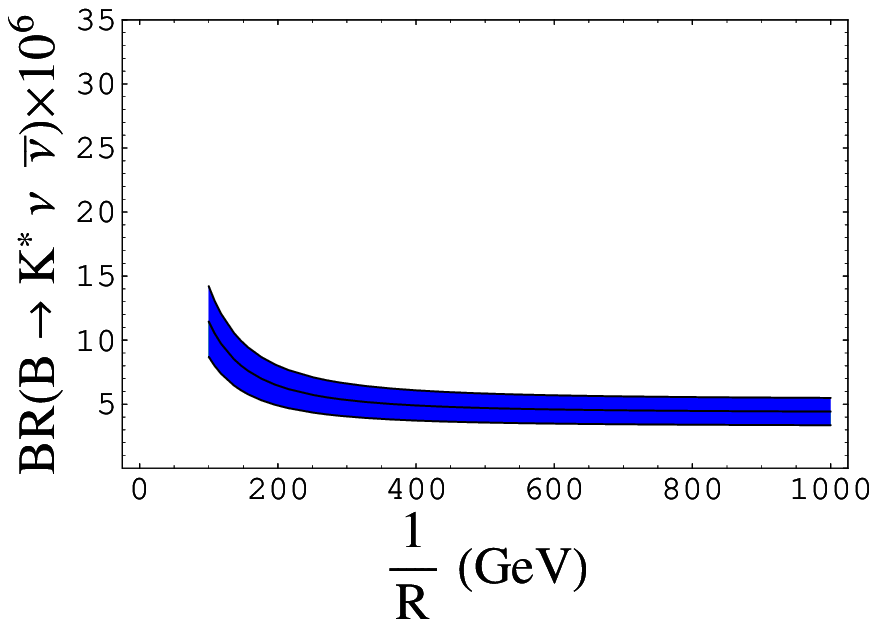}
\caption{\baselineskip=15pt Left: missing energy distribution in
$B \to K^* \nu {\bar \nu}$ for a longitudinally polarized $K^*$
(lower curves) and a transversally polarized $K^*$ (upper curves)
for set A.  The dark region corresponds to SM,
 the intermediate one  to $1/R=500$ GeV and  the light one  to $1/R=200$ GeV. Right: ${\cal B}_{B \to K^*
\nu {\bar \nu}}$ versus $1/R$, with set A of form factors.}
\vspace*{1.0cm} \label{spectrumLTnunu}
\end{figure}

For the mode $B \to K^* \nu {\bar \nu}$ it is interesting to
consider the missing energy distribution. We define $E_{miss}$ the
energy of the neutrino pair in the $B$ rest frame and consider the
dimensionless variable $x=E_{miss}/M_B$, which varies in the range
$ {1 -r \over 2} \le x \le 1-\sqrt{r} $ \noindent with
$r=M_{K^*}^2/M_B^2$. One can separately consider the missing
energy distributions for longitudinally and transversely polarized
$K^*$:
\begin{equation} {d \Gamma_L \over dx} = 3\,{ |c_L |^2 \over 24
\pi^3} { |\vec p^{~\prime}| \over M_{K^*}^2} \left [ (M_B +
M_{K^*})(M_B E'-M_{K^*}^2) A_1(q^2) - {2 M_B^2 \over M_B +
M_{K^*}} |\vec p^{~\prime}|^2 A_2(q^2) \right ]^2\,, \label{long}
\end{equation} and \begin{equation} {d \Gamma_{\pm} \over dx} = 3 { |\vec
p^{~\prime}| q^2 \over 24 \pi^3} |c_L |^2 \left |  { 2 M_B |\vec
p^{~\prime}| \over M_B + M_{K^*}} V(q^2) \mp   (M_B + M_{K^*})
A_1(q^2) \right |^2\, \label{tran} \end{equation} where $\vec
p^{~\prime}$ and $E'$ are the $K^*$ three-momentum and energy in
the $B$ meson rest frame and the sum over the three neutrino
species is understood. The missing energy distributions for
polarized $K^*$ are shown, for different values of the
compactification radius, in the left part of Fig.
\ref{spectrumLTnunu}, whereas in the right part of it, the
branching fraction is plotted as a function of the
compactification radius.

Also the decay modes $B_s \to \phi \gamma$ and $B_s \to \phi \nu
\bar \nu$  are described through the effective Hamiltonians
(\ref{hamilnu}). It is useful to relate  the branching fraction
${\cal B}_{B_s \to \phi \gamma}$ to the measured value of ${\cal
B}_{B_d \to K^{*0} \gamma}$: \begin{eqnarray}  {\cal B}_{B_s \to
\phi \gamma}= \left ({T_1^{B_s \to \phi}(0) \over T_1^{B_d \to
K^{*0}}(0)}\right)^2 \left( {M_{B_d} \over M_{B_s}} \right)^3
 \left( {M_{B_s}^2-M_\phi^2 \over M_{B_d}^2-M_{K^{*0}}^2}\right)^3 {\tau_{B_s} \over  \tau_{B_d}} {\cal B}_{B_d \to K^{*0} \gamma} \hspace{0.1cm}. \label{Bsratio}
\end{eqnarray}  This equation shows that a crucial quantity to
predict ${\cal B}_{B_s \to \phi \gamma}$ is the $SU(3)_F$ breaking
parameter $\hat{r}$ defined by $ {T_1^{B_s \to \phi}(0) \over
T_1^{B_d \to K^{*0}}(0)}=1+\hat{r}  \,\,\, $. Detailed analyses of
the range of values within which $\hat{r}$ can vary are not
available, yet. Using $\hat{r}=0.048\pm0.006$ estimated by Light
Cone Sum Rules (LCSR)  \cite{Ball:2004rg} we predict: ${\cal
B}_{B_s \to \phi \gamma}=(4.2 \pm 0.3) \times 10^{-5} \,\, . $
This prediction is compatible with the recent measurement
performed by Belle Collaboration \cite{BelleBs}:
 ${\cal
B}_{B_s \to \phi
\gamma}^{exp}=(5.7_{-1.5}^{+1.8}(stat)_{-1.1}^{+1.2}(syst))\times
10^{-5}$.

\begin{figure}[b]
 \includegraphics[width=0.35\textwidth] {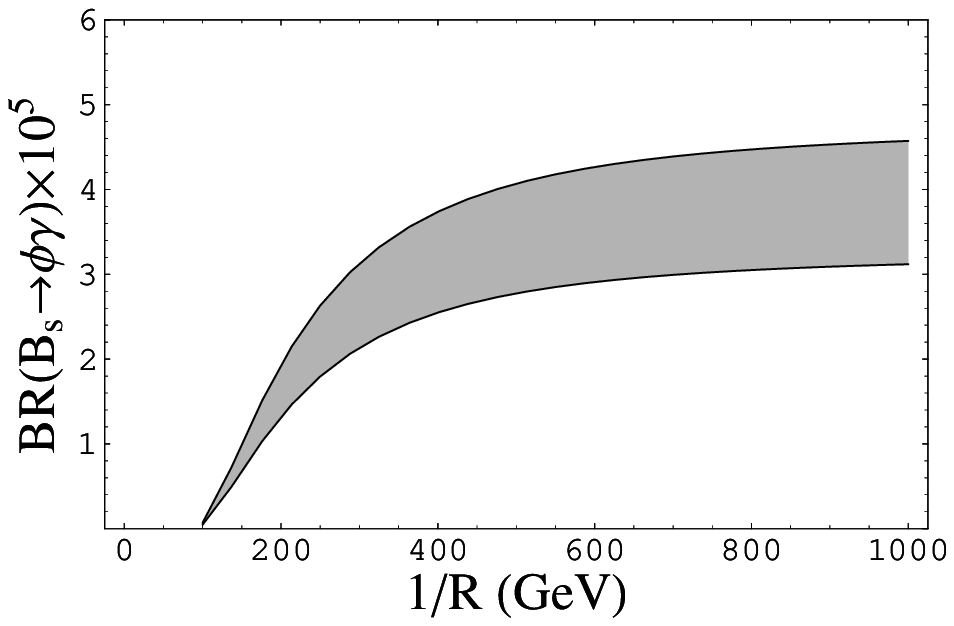}\\ \hspace{0.5cm}
 \includegraphics[width=0.35\textwidth] {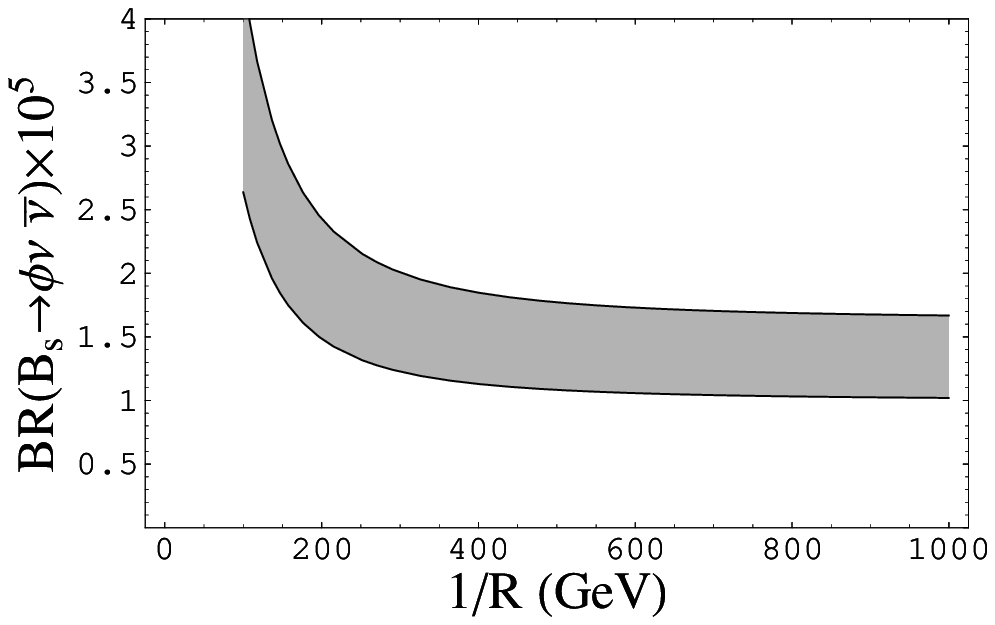}
 \caption{\baselineskip=12pt  Left: ${\cal B}(B_s \to \phi \gamma)$ {\it
vs} the compactification parameter   $1/R$. The present
 experimental upper bound is ${\cal B}_{B_s \to \phi \gamma}<12 \times 10^{-5}$  (at $90 \%$ CL).
 Right: ${\cal B}_{B_s \to \phi \nu {\bar
\nu}}$  versus  $1/R$.} \label{BsspectrumLTnunu}
\end{figure}

\noindent In the single extra dimension scenario the modification
of the Wilson coefficient $c_7$ changes  the  prediction for
${\cal B}_{B_s \to \phi \gamma}$, as shown  in Fig.
\ref{BsspectrumLTnunu} where we plot the  branching ratio versus
$1/R$.

Let us consider the mode $B_s \to\phi \nu \bar \nu$. The missing
energy distributions for polarized $\phi$ mesons have expressions
completely analogous to the case $B \to K^* \nu \bar \nu$, eqs.
(\ref{long}),(\ref{tran}). For computing their values, we have
used $B_s \to \phi$ form factors determined by LCSR
\cite{Ball:2004rg}. The distributions result to be sensitive to
the extra dimension, in particular the maximum of the
distributions is always higher than in the SM.

\noindent The SM prediction for the branching ratio is ${\cal
B}_{B_s \to \phi \nu \bar \nu}=(1.3 \pm 0.3) \times 10^{-5}$. It
suggests that this mode is within the reach of  future
experiments, although the observation of a final state involving a
neutrino-antineutrino pair is  a challenging task. The dependence
of  ${\cal B}_{B_s \to \phi \nu \bar \nu}$ on  $1/R$ is depicted
in the right part of Fig. \ref{BsspectrumLTnunu}.

\section{The decays $\Lambda_b \to \Lambda \gamma$ and $\Lambda_b \to \Lambda \nu \bar \nu$}\label{sec:lambdab}

In the case of $\Lambda_b \to \Lambda$ transitions  the hadronic
 matrix elements of the operators  $O_7$ and $O_L$ in eq. (\ref{hamilnu}) involve a larger number  of form factors.
At present, a determination  of such form factors is not
available. However, it is possible to invoke heavy quark
symmetries for the hadronic matrix elements involving    an
initial   spin=${1 \over 2}$ heavy baryon comprising a single
heavy quark Q and a final   ${1 \over 2}$ light baryon;  the heavy
quark symmetries    reduce to two the number of independent form
factors. As a matter of fact, in the infinite heavy quark limit
$m_Q \to \infty$ and  for a generic Dirac matrix $\Gamma$ one can
write \cite{Mannel:1990vg}:
\begin{equation} <\Lambda(p^\prime,s^\prime)|{\bar s} \Gamma b
|\Lambda_b(p,s)>  = {\bar u}_\Lambda(p^\prime,s^\prime) \big\{
F_1(p^\prime \cdot v) + \spur{v} F_2  (p^\prime \cdot v) \big\}
\Gamma u_{\Lambda_b}(v,s) \nonumber \\  \label{hqrelations}
\end{equation} where $u_\Lambda$ and $u_{\Lambda_b}$ are the $\Lambda$  and
$\Lambda_b$ spinors, and $v={p \over M_{\Lambda_b}}$  is the
$\Lambda_b$ four-velocity. The form factors $F_{1,2}$ depend on
$\displaystyle{p^\prime \cdot v= {M^2_{\Lambda_b}+M^2_\Lambda-q^2
\over 2 M_{\Lambda_b}}}$ (for convenience we instead consider them
as functions of $q^2$ through this relation).

A determination of  $F_1$ and  $F_2$ has been  obtained by
three-point QCD sum rules in the $m_Q \to \infty$ limit
\cite{Huang:1998ek}. In  the following  we  use the expressions
for the functions $F_1$, $F_2$ obtained by  updating some of the
parameters used  in \cite{Huang:1998ek}:

 \begin{equation}
 F_{1,2}(q^2)={F_{1,2}(0) \over 1+a_{1,2} \, q^2+b_{1,2} \, q^4} \label{fit-ff-huang}\end{equation}
with $F_1(0) = 0.322 \pm 0.015$,     $a_1=-0.0187$ GeV$^{-2}$,
$b_1=-1.6\times 10^{-4}$ GeV$^{-4}$,  and $F_2(0) =-0.054 \pm
0.020$,    $a_2=-0.069$ GeV$^{-2}$,   $b_2=1.5\times 10^{-3}$
GeV$^{-4}$.

\noindent The knowledge of  $\Lambda_b$ form factors deserves a
substantial  improvement;  in the meanwhile, we use in our
analysis the form factors in (\ref{fit-ff-huang}) stressing that
the uncertainties attached to the various predictions  only take
into account  the errors  of  the parameters of the model chosen
for   $F_1$ and $F_2$.

\begin{figure}[ht]
\includegraphics[width=0.35\textwidth]  {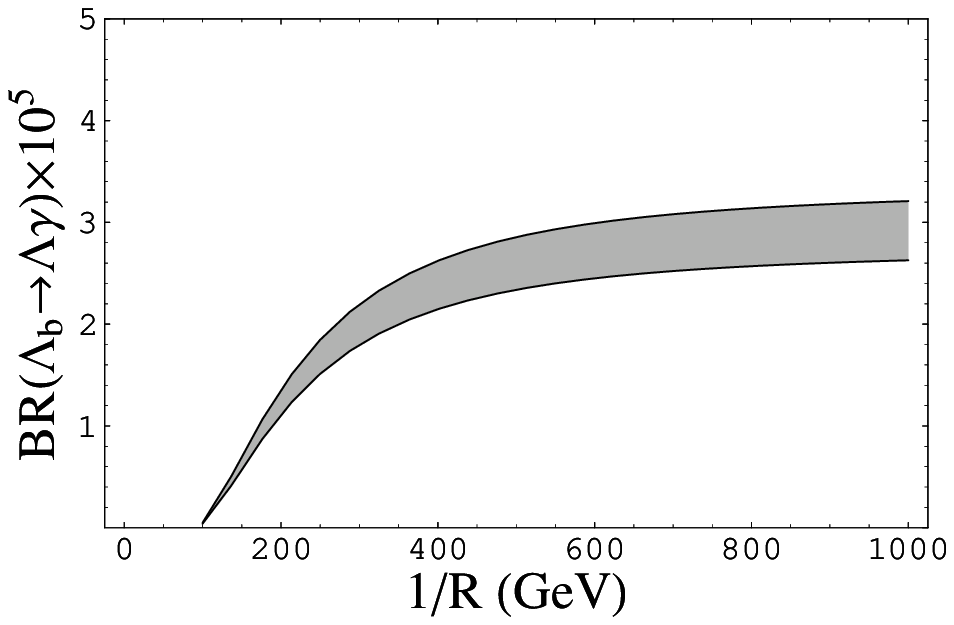} \hspace{0.5cm}
 \includegraphics[width=0.35\textwidth] {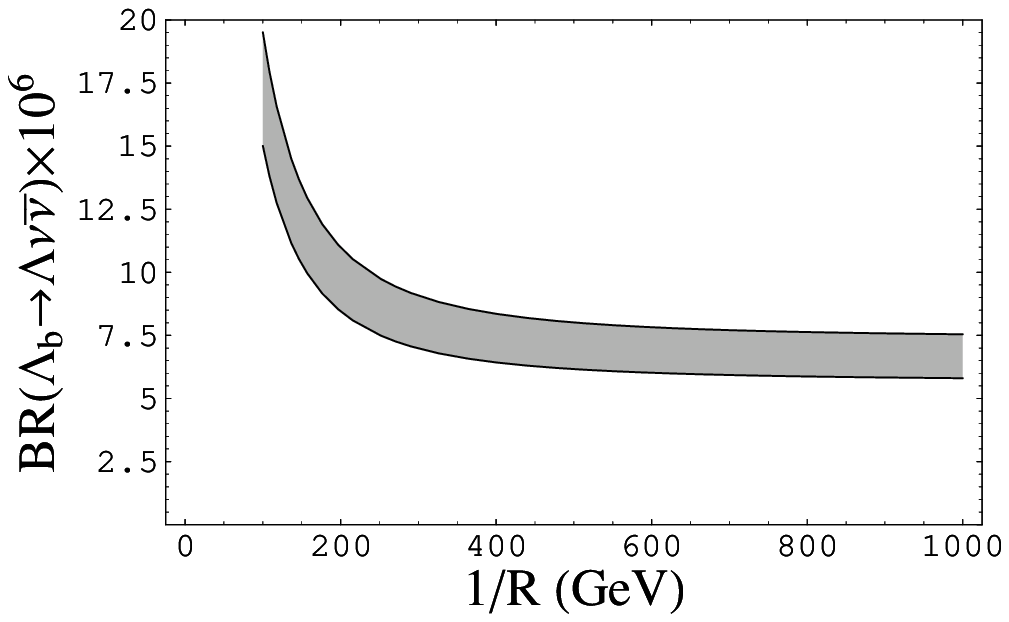}
\caption{\baselineskip=15pt ${\cal B}_{\Lambda_b \to \Lambda
\gamma}$ (left) and ${\cal B}_{\Lambda_b \to \Lambda \nu \bar
\nu}$ (right) plotted as function of $1/R$.  The uncertainty shown
by the dark band is mainly due to the errors on the form
 factors of the  model (\ref{fit-ff-huang}).} \vspace*{1.0cm}
\label{brllnunuhuang}
\end{figure}

The $\Lambda_b \to \Lambda \gamma$ branching ratio can be related
to that of  $B_d \to K^{*0} \gamma$  through the expression:
\begin{eqnarray}
 {\cal B}_{\Lambda_b \to \Lambda \gamma}= \left ({F_1(0) \over
T_1^{B_d \to K^{*0}}(0)}\right)^2 \left(1+{M_\Lambda \over
M_{\Lambda_b}} {F_2(0) \over F_1(0)} \right)^2
  \left( {M_{B_d} \over M_{\Lambda_b}}  {M_{\Lambda_b}^2-M_\Lambda^2 \over M_{B_d}^2-M_{K^{*0}}^2}\right)^3 {\tau_{\Lambda_b} \over  4 \tau_{B_d}} {\cal B}_{B_d \to K^{*0} \gamma} \,\,\,
 \nonumber   \end{eqnarray}
Using the ratio of form factors obtained from (\ref{fit-ff-huang})
and the form factor $T_1$ in \cite{Ball:2004rg},
 we predict:  ${\cal B}_{\Lambda_b \to \Lambda \gamma}=(3.4 \pm 0.7) \times
 10^{-5}$. As for  the effect of the extra dimension in
modifying the decay rate, in Fig. \ref{brllnunuhuang} we show how
${\cal B}_{\Lambda_b \to \Lambda \gamma}$ depends on $1/R$.

For the mode $\Lambda_b \to \Lambda \nu \bar \nu$, in Fig.
\ref{brllnunuhuang} (right part) we plot the dependence of the
branching ratio on $1/R$. The SM value is expected to be $ {\cal
B}_{\Lambda_b \to \Lambda \nu \bar \nu}=(6.5 \pm 0.9) \times
10^{-6} $.

\noindent These results suggest that these processes  are within
the reach of LHC experiments.

\section{Conclusions} \label{sec:concl}

We have studied how  a single universal extra dimension could have
an impact on  several  loop induced $B_{d,s}$ and $\Lambda_b$
decays. For the $B \to K^* \gamma$ mode, the uncertainty related
to the form factors does not obscure the sensitivity to the
compactification radius. From the comparison between predictions
and experimental data we obtain a lower bound of $1/R \geq
300-400$ GeV. Then, we have found that hadronic uncertainties are
not large in case of $B_s$ decays, whereas for $\Lambda_b$ the
situation is more uncertain. These results can be useful for the
Physics programs at the hadron colliders.


\begin{theacknowledgments}
I acknowledge partial support from the EU contract No.
MRTN-CT-2006-035482, "FLAVIAnet".

\end{theacknowledgments}

\bibliographystyle{aipproc}

\begin{thebibliography}{9}

\bibitem{rev}
  I.~Antoniadis,
  Phys.\ Lett.\ B {\bf 246}, 377 (1990);
  K.~R.~Dienes, E.~Dudas and T.~Gherghetta,
  Phys.\ Lett.\ B {\bf 436}, 55 (1998);
  N.~Arkani-Hamed and M.~Schmaltz,
  Phys.\ Rev.\ D {\bf 61}, 033005 (2000).

\bibitem{ACD}
  T.~Appelquist, H.~C.~Cheng and B.~A.~Dobrescu,
  Phys.\ Rev.\ D {\bf 64}, 035002 (2001).

  \bibitem{Agashe:2001xt}
  K.~Agashe, N.~G.~Deshpande and G.~H.~Wu,
  Phys.\ Lett.\ B {\bf 514}, 309 (2001).

  \bibitem{Buras:2002ej}
  A.~J.~Buras, M.~Spranger and A.~Weiler,
  Nucl.\ Phys.\ B {\bf 660}, 225 (2003);
  A.~J.~Buras, A.~Poschenrieder, M.~Spranger and A.~Weiler,
  Nucl.\ Phys.\ B {\bf 678}, 455 (2004).

\bibitem{Haisch:2007vb}
  U.~Haisch and A.~Weiler,
  Phys.\ Rev.\  D {\bf 76}, 034014 (2007).

   \bibitem{noi}
    P.~Colangelo, F.~De Fazio, R.~Ferrandes and T.~N.~Pham,
  Phys.\ Rev.\ D {\bf 73}, 115006 (2006);
  Phys.\ Rev.\  D {\bf 74}, 115006 (2006);
  arXiv:0709.2817 [hep-ph].

  \bibitem{inami}
T.~Inami and C.~S.~Lim,
  Prog.\ Theor.\ Phys.\  {\bf 65}, 297 (1981)
  [Erratum-ibid.\  {\bf 65}, 1772 (1981)].

\bibitem{buchalla}
G.~Buchalla and A.~J.~Buras,
  Nucl.\ Phys.\ B {\bf 400}, 225 (1993);
G.~Buchalla, A.~J.~Buras and M.~E.~Lautenbacher,
  Rev.\ Mod.\ Phys.\  {\bf 68}, 1125 (1996).

  \bibitem{Buchalla:1998ba}
  G.~Buchalla and A.~J.~Buras,
  Nucl.\ Phys.\ B {\bf 548}, 309 (1999).

\bibitem{Colangelo:1995jv}
  P.~Colangelo, F.~De Fazio, P.~Santorelli and E.~Scrimieri,
  Phys.\ Rev.\ D {\bf 53}, 3672 (1996)
  [Erratum-ibid.\ D {\bf 57}, 3186 (1998)].

  \bibitem{Ball:2004rg}
  P.~Ball and R.~Zwicky,
  Phys.\ Rev.\ D {\bf 71}, 014015 (2005);
  Phys.\ Rev.\ D {\bf 71}, 014029 (2005).



\bibitem{Nakao:2004th}
  M.~Nakao {\it et al.}  [BELLE Collaboration],
  Phys.\ Rev.\ D {\bf 69}, 112001 (2004);
  B.~Aubert {\it et al.}  [BABAR Collaboration],
  Phys.\ Rev.\ D {\bf 70}, 112006 (2004).

\bibitem{BelleBs}
A. Drutskoy, arXiv:0710.1647 [hep-ex].


  \bibitem{PDG}
  W.~M.~Yao {\it et al.}  [Particle Data Group],
  J.\ Phys.\ G {\bf 33}, 1 (2006).

\bibitem{Mannel:1990vg}
  T.~Mannel, W.~Roberts and Z.~Ryzak,
  Nucl.\ Phys.\  B {\bf 355}, 38 (1991).

\bibitem{Huang:1998ek}
  C.~S.~Huang and H.~G.~Yan,
  Phys.\ Rev.\  D {\bf 59}, 114022 (1999)
  [Erratum-ibid.\  D {\bf 61}, 039901 (2000)].


\end{thebibliography}

\IfFileExists{\jobname.bbl}{}
 {\typeout{}
  \typeout{******************************************}
  \typeout{** Please run "bibtex \jobname" to optain}
  \typeout{** the bibliography and then re-run LaTeX}
  \typeout{** twice to fix the references!}
  \typeout{******************************************}
  \typeout{}
 }

\end{document}